\documentclass[10pt,amsmath,amssymb,nofootinbib,aps]{revtex4} 
\usepackage{bm}
\usepackage{graphics,calc,epsfig,pstricks,bbm}
\usepackage{amsmath,amssymb,amsfonts}

\newcommand{\be}{\begin{equation}}
\newcommand{\ee}{\end{equation}}
\newcommand{\bea}{\begin{eqnarray}}
\newcommand{\eea}{\end{eqnarray}}

\begin{document}
\title{Localization and diffusion in polymer quantum field theory}

\author{Michele Arzano}
\email{michele.arzano@roma1.infn.it}
\affiliation{Dipartimento di Fisica and INFN,\\ ``Sapienza" University of Rome,\\ P.le A. Moro 2, 00185 Roma, EU }
\author{Marco Letizia}
\email{marco.letizia@sissa.it}
\affiliation{\textit{SISSA},\\ via Bonomea 265, 34136 Trieste, Italy }

\begin{abstract}
\begin{center}
{\bf Abstract}\\
\end{center}
Polymer quantization is a non-standard approach to quantizing a classical system inspired by background independent approaches to quantum gravity such as loop quantum gravity.  When applied to field theory it introduces a characteristic {\it polymer scale} at the level of the fields classical configuration space.  Compared with models with space-time discreteness or non-commutativity this is an alternative way in which a characteristic scale can be introduced in a field theoretic context.  Motivated by this comparison we study here localization and diffusion properties associated with polymer field observables and dispersion relation in order to shed some light on the novel physical features introduced by polymer quantization.  While localization processes seems to be only mildly affected by polymer effects, we find that polymer diffusion differs significantly from the ``dimensional reduction" picture emerging in other Planck-scale models beyond local quantum field theory. 
\end{abstract}

\maketitle
\noindent

\section{Introduction}
The fate of the basic structures of local quantum field theory is one of the main open questions in the attempts to include gravity in the quantum picture of forces and matter.  According to common wisdom the gravitational coupling constant in a quantum context introduces a characteristic length scale (the Planck length $l_p=\sqrt{\hbar G/c^3}$) where the description of space-time will likely be quite far from the familiar picture of a smooth manifold.  The analysis of the propagation of particles in such {\it quantum} space-times could dictate departures from certain fundamental notions like locality and local Lorentz invariance which are among the building blocks of ordinary quantum field theory, the language at the basis of our understanding of high energy particle physics.\\
An immediate, and rather crude, way such quantum gravity scale could enter a field theoretic setting is through violations of Lorentz invariance.  Departures from Lorentz invariance have indeed been suggested in several quantum gravity frameworks contemplating space-time discreteness or ``non-commutativity" set by the Planck length  \cite{Gambini:1998it,Alfaro:1999wd,Kostelecky:1989jw,Douglas:2001ba}.  However there seems to be no quantum gravity approach in which theoretical evidence for Planck scale violations of Lorentz invariance is conclusive even though experimental searches for high energy departures from Lorentz symmetry have been very active in the past decade \cite{lorentztests}.\\
Another way to introduce in a {\it structural} way a characteristic quantum gravity energy scale in quantum field theory is via {\it curvature} of momentum space.  The most studied example in this context is the case of momenta defined on a non-abelian Lie group which carries a transitive action of the Lorentz group.  In such models relativistic symmetries are preserved but {\it deformed} \cite{Lukierski:1992dt,AmelinoCamelia:2000mn,AmelinoCamelia:2002ar,Arzano:2010jw}.  The corresponding field theory on coordinate space is a non-commutative field theory defined on a Lie algebra.  Even though the most studied examples of such models were introduced in the case of four-dimensional space-times adopting as momentum space a sub-manifold of de Sitter space (see e.g. \cite{KowalskiGlikman:2003we,deSitter1,deSitter2} and references therein) the closest connection with gravity is realized for point particles coupled to Einstein gravity in 2+1 dimensions. In this context the momentum space is given by the three-dimensional Lorentz group \cite{Matschull:1997du,Arzano:2014ppa} and relativistic symmetries are described by a {\it quantum deformation} of the Poincar\'e group known as the ``quantum double" of the Lorentz group \cite{Bais:1998yn,Bais:2002ye}.\\
Rather than introducing a quantum gravity scale at the momentum space level one could look at models where the space of {\it momenta conjugate} to a field contains an intrinsic scale.  This would amount to introduce the scale directly in the structure of the phase space of the classical field.  Field theoretic models of this sort have been actually studied in recent years in connection with loop quantum gravity (LQG) \cite{lqg,Ashtekar:2012np,Perez:2012wv} and are based on a non-standard, {\it polymer}, quantization for the fields \cite{pqft1,pqft2}.   Such quantization is performed using fundamental operators similar to the basic operators of LQG, \textit{holonomies} and \textit{fluxes}, for which the usual notion of conjugate momentum is not well defined.  The associated Hilbert space is constructed using an inner product which does not rely on any background metric or space-time and thus these models can be seen as implementing a ``background independent quantization" in the same spirit as in the LQG program.\\
While in the first two examples mentioned above, the Planck length or energy set the scale for discreteness of space-time or provide a characteristic momentum, the ``polymer" scale enters directly at the level of the field space and thus the resulting space-time picture is less clear.  A natural question one might ask for example is whether the polymer scale sets any kind of limitation on the measurement of lengths or momenta or if it affects the way a propagating particle {\it sees} the underlying space-time.  In this work we provide the first exploration of these questions in the context of the approach to polymer quantum field theory proposed and studied in \cite{pqft1,pqft2}.  In particular we study two different procedures of localization using quantum field theoretic tools, one in which the the restriction to a given region is obtained through boundary conditions on the field modes and the other where localization is imposed directly at the level of quantum states.  We further analyze a diffusion process using a polymer corrected Laplacian and calculate the associated spectral dimension. The resulting picture shows that polymer corrections affect in a rather mild way the usual procedures of localization while the running of the polymer spectral dimension with the diffusion scale reveals {\it superdiffusion} at the polymer scale and the absence of any ``dimensional reduction" in the deep UV, a common feature of many quantum gravity scenarios \cite{Ambjorn:2005db,Lauscher:2005qz,Horava:2009if,Benedetti:2008gu,Modesto:2008jz,Carlip:2009kf,Calcagni:2010pa,Sotiriou:2011mu,Alesci:2011cg,Amelino-Camelia:2013cfa,Arzano:2014jfa}.\\
The paper is organized as follows: in Section II we start by reviewing polymer quantization in field theory following the prescriptions given in \cite{pqft1} focussing our attention on the non-standard behaviour of Fock operators, the definition of the vacuum state and some relevant expectation values. In section III we introduce two different approaches to localization in quantum field theory, one based on boundary conditions on field operators the other on ``localized" one-particle states. In Section IV we propose an extension of these analyses to the context of polymer quantum field theory. In particular we study the localization procedure as the region of the local volume (the size of the measurement apparatus) approaches the characteristic, polymer, scale of the theory and the behaviour of a sharply localized polymer quantum state. In Section V we analyze diffusion properties in a polymer setting, first defining a polymer trace of the heat kernel and then calculating the associated spectral dimension. We conclude by discussing our results and suggesting possible developments.

\section{The setting: polymer quantum fields}
In this Section we review the approach to polymer quantization for a scalar field proposed in \cite{pqft1}.  Such approach relies on a modified canonical field quantization and definition of Fock space which is well suited for our purposes. In 3+1 dimensional Minkowski space the Hamiltonian of a real scalar field reads
\begin{equation}
H=\int_V \frac{1}{2}\bigg( \pi^2+\partial_a\phi\partial^a\phi+m^2\phi^2 \bigg) d^3x\, ,
\end{equation}
where the canonical phase space variables are the field and its conjugated momentum $(\phi(\vec{x},t),\pi(\vec{x},t))$.  The first step towards polymer quantization is a re-definition of such basic phase space variables which introduces a {\it polymer} scale.  The new phase space variables $(\Phi(\vec{x},t),U_{\lambda}(\vec{x},t))$ are defined by \cite{pqft1,pqft2}
\begin{equation}\label{pol_variables}
\begin{split}
&\Phi(\vec{x},t)=\frac{1}{V}\int_V f(\vec{x}-\vec{x}')\phi(\vec{x}',t) d^3x',\\
&U_{\lambda}(\vec{x},t)=e^{\imath\lambda \pi(\vec{x},t)}\, .
\end{split}
\end{equation}
Here the function $f$ is a real valued test function, $\lambda$ is a real constant with the dimension of a length squared (in three spatial dimension and natural units) and $V$ is the volume of 3-space. The equal time canonical Poisson bracket is given by
\begin{equation}
\big\{ \Phi_f (\vec{x},t),U_\lambda (\vec{x}',t)\big\}_{\phi,\pi}=\imath\frac{\lambda}{V}f(\vec{x}-\vec{x}')U_\lambda(\vec{x}',t).
\end{equation} 
The Hilbert space on which the quantum counterparts of $\Phi(\vec{x},t)$ and $U_{\lambda}(\vec{x},t)$ act is constructed from basis states given by
\begin{equation}
|\mu_1,...,\mu_N\rangle\,,
\end{equation}
where $\{\mu_i|i=1,...,N\}$ are real numbers, each of which is the value of the field at the i-th point of a spatial {\it graph} $\{x_i|i=1,...,N\}$, i.e. a discrete set of points scattered in space.  The inner product on such Hilbert space is given by
\begin{equation}
\langle \nu_1,...,\nu_N|\mu_1,...,\mu_N\rangle=\delta_{\nu_1,\mu_1}\cdot...\cdot\delta_{\nu_N,\mu_N},
\end{equation}
which can be seen as a {\it background independent} inner product since it makes no reference to the background metric. This is to contrast with the usual Fock space quantization based on the metric dependent Klein-Gordon inner product.  Notice here that there is no particular scale associated with the graph unless one specifically chooses a uniform lattice.\\
It is customary to take $f$ to be a Gaussian sharply peaked at a point $x_j$ and we set $\Phi_f(\vec{x}_j,t)\equiv\Phi_j$ and $U(\vec{x}_j,t)\equiv U_j$.  One then promotes the basic variables to operators on the Hilbert space by imposing the commutators\footnote{Note that, choosing $f(\vec{x}-\vec{x}')=e^{-\frac{(\vec{x}-\vec{x}')^2}{\sigma^2}}$, we have $f(\vec{x}-\vec{x}')\approx 1$ for $(\vec{x}-\vec{x}')^2\ll\sigma^2$ and $f(\vec{x}-\vec{x}')\approx 0$ otherwise.}
\begin{equation}
\begin{split}
&[\tilde{\Phi}_j,\tilde{U}_l]=-\frac{\lambda}{V}\delta_{j,l}\tilde{U}_l,\\
&[\tilde{\Phi}_j,\tilde{U}^{\dagger}_l]=\frac{\lambda}{V}\delta_{j,l}\tilde{U}^{\dagger}_l\,;
\end{split}
\end{equation}
where $\delta_{jl}\equiv\delta_{x_j,x_l}$. The action of the basis phase space functions is given by\footnote{We will drop the tilde over the operators from now on.}
\begin{equation}\label{action_operators}
\begin{split}
&\Phi_j|\mu_1...\mu_N\rangle=\mu_j|\mu_1...\mu_N\rangle, \\
&U_j|\mu_1...\mu_N\rangle=|\mu_1...\mu_j-\frac{\lambda}{V}...\mu_N\rangle,\\
&U^{\dagger}_j|\mu_1...\mu_N\rangle=|\mu_1...\mu_j+\frac{\lambda}{V}...\mu_N\rangle.\\
\end{split}
\end{equation}
Although the space has a discrete character $\lambda$ does not refer to the discreteness of space, but, as anticipated in the Introduction, is a fundamental scale in field configuration space.  Basis states in the Hilbert space are now orthogonal when associated with different graphs or with different excitations at a point.  More importantly since the inner product on basis states is given by a discrete delta the generator of infinitesimal shift in configuration space i.e., the momentum operator $\pi(\vec{x},t)$, is not well defined and one has to resort to a \textit{polymerized momentum operator} 
\begin{equation}\label{mom_reg}
\pi_j=\frac{1}{2\imath\lambda}(U_j-U_j^{\dagger})\, ,
\end{equation}
which has a well defined action and accordingly the canonical commutator gets modified from that in usual field theory to
\begin{equation}
\big[\Phi_j,\pi_l\big]=\frac{\imath}{2V}\delta_{j,l}\big(U_j+U^{\dagger}_j\big).
\end{equation}
In analogy with ordinary QFT one can introduce creation and annihilation operators and construct a suitable Fock space which, as we will see in detail below, exhibits various non-trivial features.  We can proceed in analogy with the standard case where one inverts the Fourier expansion for the field and the momentum conjugate to get
\begin{equation}
\begin{split}
&a_{\vec{k}}=\imath\int_V f^{\dagger}_{\vec{k}} (\vec{x},t) \bigg(\pi-\imath\omega_{\vec{k}}\phi\bigg) d^3x,\\
&a^{\dagger}_{\vec{k}}=-\imath\int_V f_{\vec{k}} (\vec{x},t) \bigg(\pi+\imath\omega_{\vec{k}}\phi\bigg) d^3x,\\
\end{split}
\end{equation}
where $f_{\vec{k}}(\vec{x},t)=\frac{e^{-\imath\omega_{\vec{k}}t+{\imath\vec{k}\cdot\vec{x}}}}{\sqrt{2\omega_{\vec{k}}V}}$ are the usual plane waves with $\vec{k}=2\pi\mathbb{Z}^3/L$, $L^3=V$.  Following \cite{pqft1} we can then use the field and momentum operators introduced above to define the corresponding creation and annihilation operators in the polymer context. We take a generic graph $\{\vec{x}_j|j=1,...,N\}$ and define the operators 
\begin{equation}\label{pol_operators}
\begin{split}
&A_{\vec{k}}=\frac{\imath V}{\sqrt{N}}\sum_{j=1}^N f^{\dagger}_{\vec{k}j}\big( \pi_j-\imath\omega_{\vec{k}}\Phi_{j}\big),\\
&A_{\vec{k}}^{\dagger}=-\frac{\imath V}{\sqrt{N}}\sum_{j=1}^N f_{\vec{k}j}\big( \pi_j+\imath\omega_{\vec{k}}\Phi_{j}\big),\\
\end{split}
\end{equation}
where we used the shorthand notation $f_{\vec{k}j}=f_{\vec{k}}(x_j)$.\\
Using the {\it polymer} creation and annihilation operators just defined we can introduce a notion of Fock space.  To do so we need a candidate vacuum state.  One possibility is to define the latter in such a way that expectation values of polymer observables reduce to vacuum expectation values of ordinary operators in QFT in the limit when the polymer scale vanishes.  Adopting the picture of a quantum field described in terms of an infinite collection of harmonic oscillators, one for every spatial point $x$ and mode $k$, the polymer vacuum can be defined in terms of a product of vacua of simple harmonic oscillators, one for each point of the graph. In practice we start with the following state
\begin{equation}
|G_{\vec{k}}\rangle_j=\frac{1}{\pi^{1/4}}\sum_\mu e^{-\frac{\omega_{\vec{k}}V\mu^2}{2}}|\mu\rangle_j.
\end{equation}
where $V$ is the same volume appearing in (\ref{pol_variables}). This ground state of an ensemble of oscillators of frequency $\omega_{\vec{k}}$ at the point $\vec{x}_j$. It is easily checked that this state is normalized to $1$ if we evaluate the sum by means of the continuum limit
\begin{equation}\label{vacuum_gauss}
\sum_\mu \longrightarrow \sqrt{\omega_{\vec{k}}V}\int_\mathbb{R} d\mu\,.
\end{equation}
The polymer vacuum for a single mode $\vec{k}$ is then given by
\begin{equation}\label{pol_vacuum}
|0_{\vec{k}}\rangle=\bigotimes_{j=1}^N|G_{\vec{k}}\rangle_j,
\end{equation}
and thus we can define the ``full" polymer vacuum state as
\begin{equation}\label{fullvacuum}
|0\rangle=\bigotimes_{\vec{k}}|0_{\vec{k}}\rangle.
\end{equation}
Finally one can define, in analogy with ordinary field theory, a multi-particle state as
\begin{equation}\label{multip}
|n_{\vec{k}}\rangle=\frac{1}{\sqrt{\langle n_{\vec{k}}|n_{\vec{k}}\rangle}}\big(A^{\dagger}_{\vec{k}}\big)^n |0_{\vec{k}}\rangle.
\end{equation}
After this flash review of the basic ingredients of polymer field quantization, we elaborate and further develop some results first derived in \cite{pqft1}, in a way which will be useful for our analysis. In particular we will study the behaviour of the expectation values of simple observables. To simplify the notation we introduce the dimensionless parameter $\gamma_{\vec{k}}=\omega_{\vec{k}}\lambda^2/V$. We start from the vacuum expectation value of the number operator $N_{\vec{k}}=A^{\dagger}_{\vec{k}}A_{\vec{k}}$. Using (\ref{action_operators}),(\ref{mom_reg}),(\ref{pol_operators}) and the definition of polymer vacuum state (\ref{fullvacuum}) one obtains
\begin{equation}\label{number_on_vacuum}
\langle 0_{\vec{k}}|N_{\vec{k}}|0_{\vec{k}}\rangle=\frac{1}{4}-\frac{e^{-\gamma_{\vec{k}}/4}}{2}+\frac{1-e^{\gamma_{\vec{k}}}}{4\gamma_{\vec{k}}}.
\end{equation}
The non-vanishing right-hand side of (\ref{number_on_vacuum}) reveals the unusual feature of the polymer vacuum state of not being annihilated by $A_{\vec{k}}$ in contrast with ordinary quantum field theory where this is a defining property of the vacuum state.\\
As a next step we introduce the Hamiltonian in analogy with ordinary QFT 
\begin{equation}\label{pol_ham}
H=\sum_{\vec{k}}\omega_k \bigg(A_{\vec{k}}^{\dagger}A_{\vec{k}}+\frac{1}{2}\big[A_{\vec{k}},A_{\vec{k}}^{\dagger} \big] \bigg)\,,
\end{equation}
and calculate its polymer vacuum expectation value
\begin{equation}\label{energy_vacuum}
\langle 0|H|0\rangle=\frac{1}{4}\sum_{\vec{k}} \omega_{\vec{k}}\bigg( 1+\frac{1-e^{-\gamma_{\vec{k}}}}{\gamma_{\vec{k}}}\bigg).
\end{equation}
We see that the expectation value above gives us a polymer corrected energy associated with the mode $\vec{k}$
\begin{equation}
E_{\vec{k}}=\frac{1}{2}\omega_{\vec{k}}\bigg(1+\frac{1-e^{-\gamma_{\vec{k}}}}{\gamma_{\vec{k}}}\bigg)\,,
\end{equation}
which indicates that polymer effects lead to a {\it modified dispersion relation}. Expanding the expression above for $\gamma_{\vec{k}}\ll1$ we get
\begin{equation}\label{disp_rel}
E^2_{\vec{k}}\simeq m^2+|\vec{k}|^2-\frac{|\vec{k}|^3\ell}{2}
\end{equation}
where we introduced a characteristic polymer length scale $\ell=\frac{\lambda^2}{V}$ for the leading order polymer correction to the dispersion relation.\\
Let us now turn to ``one particle states". We look again at the expectation value of the Hamiltonian (\ref{pol_ham}). For the one particle state defined in (\ref{multip}) we want to calculate the ``normalized" one-particle energy via
\begin{equation}
\begin{split}
&\langle \vec{p}|H|\vec{p}\rangle=\frac{1}{\langle \vec{p}|\vec{p}\rangle}\langle 0|A_{\vec{p}}\bigg[\sum_{\vec{k}}\omega_{\vec{k}} \bigg(A_{\vec{k}}^{\dagger}A_{\vec{k}}+\frac{1}{2}\big[A_{\vec{k}},A_{\vec{k}}^{\dagger} \big] \bigg)\bigg]A^{\dagger}_{\vec{p}}|0\rangle=\\
&=\frac{1}{2\langle \vec{p}|\vec{p}\rangle}\langle 0|A_{\vec{p}}\bigg[\omega_{\vec{p}} \bigg(A_{\vec{p}}^{\dagger}A_{\vec{p}}+A_{p}A^{\dagger}_{\vec{p}}\bigg)+\sum_{\vec{k}\neq \vec{p}}\omega_{\vec{k}} \bigg(A_{\vec{k}}^{\dagger}A_{\vec{k}}+A_{\vec{k}}A^{\dagger}_{\vec{k}}\bigg)\bigg]A^{\dagger}_{\vec{p}}|0\rangle.
\end{split}
\end{equation}
As a first step we evaluate the ``one-particle" expectation value of the number operator
\begin{equation}
\begin{split}
&\langle \vec{p}|N_{\vec{k}}|\vec{p}\rangle=\frac{1}{\langle \vec{p}|\vec{p}\rangle}\langle 0|A_{\vec{p}}(A_{\vec{k}}^{\dagger}A_{\vec{k}})A^{\dagger}_{\vec{p}}|0\rangle=\\
&=\frac{1}{16 \gamma_{\vec{p}} \bigg(2 e^{\frac{3}{4}\gamma_{\vec{p}}} \gamma_{\vec{p}}+e^{\gamma_{\vec{p}}} x+e^{\gamma_{\vec{p}}}-1\bigg)}\bigg(4 e^{\frac{3}{4}\gamma_{\vec{p}}} \gamma_{\vec{p}}^3+16 e^{\frac{3}{4}\gamma_{\vec{p}}} \gamma_{\vec{p}}^2+16 e^{\gamma_{\vec{p}}} \gamma_{\vec{p}}^2-4 \gamma_{\vec{p}}^2+\\
&-4 e^{-\frac{5}{4}\gamma_{\vec{p}}} \gamma_{\vec{p}}-8 e^{-\frac{\gamma_{\vec{p}}}{4}} \gamma_{\vec{p}}+12 e^{\frac{3}{4}\gamma_{\vec{p}}} \gamma_{\vec{p}}+8 e^{\gamma_{\vec{p}}} \gamma_{\vec{p}}-8 \gamma_{\vec{p}}+e^{-3 \gamma_{\vec{p}}}+6 e^{-\gamma_{\vec{p}}}+9 e^{\gamma_{\vec{p}}}-16\bigg)\,.
\end{split}
\end{equation}
\begin{figure}[htbp]
\begin{center}
\includegraphics[width=9cm, height= 6cm]{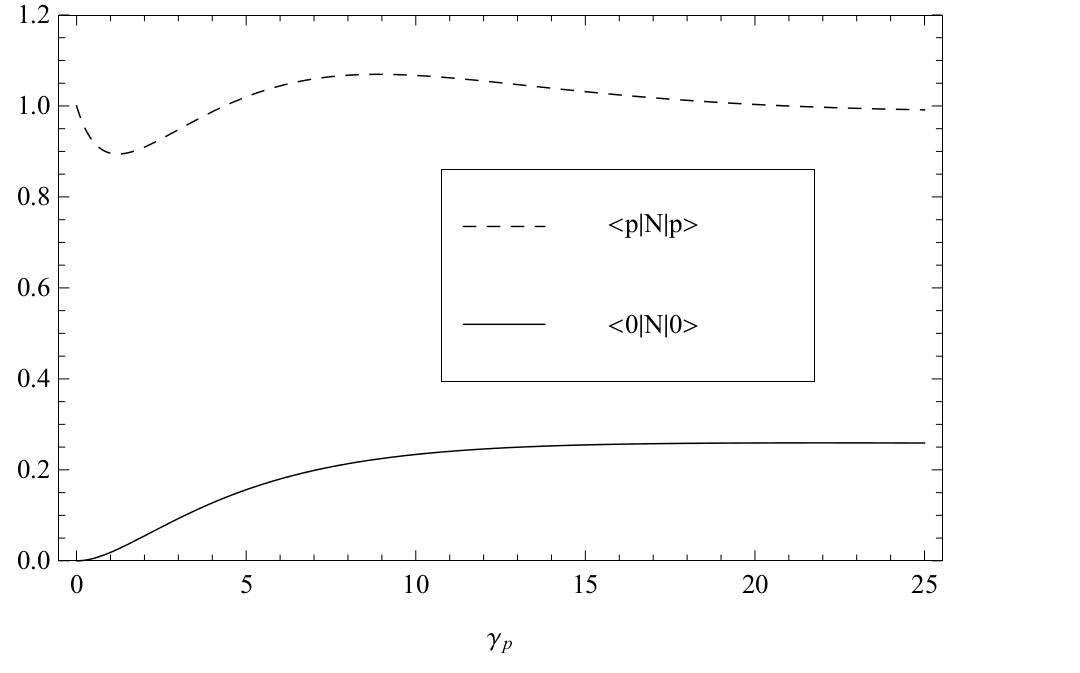}
\caption{Expectation value of the number operator in the vacuum state and in a one particle state as a function of $\gamma_{\vec{p}}$.}
\label{number}
\end{center}
\end{figure}
In Figure \ref{number} we compare the expectation value of the number operator in vacuum and in a one-particle state. We can see that the expectation value of the number operator gives fractional results, for $\gamma_{\vec{k}}\neq 0$, in both cases. This is an interesting consequence of the presence of the polymer scale $\lambda$ in the definition of the conjugate momentum $\pi$. It is an indicator that with with our choice of vacuum state and creation and annihilation operators the interpretation of $|\vec{p}\rangle$ as a ``one-polymer particle" state is somewhat problematic.\\
As a next step we look at the energy carried by one-particle states. To do so we evaluate the expectation value of the Hamiltonian in such states and subtract the vacuum energy contribution
\begin{equation}\label{oneparticle_energy_polymer}
\begin{split}
&\langle \vec{p}|H|\vec{p}\rangle-\langle 0|H|0\rangle=\\
&=\frac{\omega_{\vec{p}}}{16\gamma_{\vec{p}} \bigg(-e^{-\gamma_{\vec{p}}}+1+2e^{-\frac{1}{4}\gamma_{\vec{p}}}\gamma_{\vec{p}}+\gamma_{\vec{p}} \bigg) }\bigg(5+12\gamma_{\vec{p}}-6e^{-\frac{9}{4}\gamma_{\vec{p}}}\gamma_{\vec{p}}+10e^{-\frac{5}{4}\gamma_{\vec{p}}}\gamma_{\vec{p}}+\\
&-4e^{-\frac{5}{4}\gamma_{\vec{p}}}\gamma_{\vec{p}}+4e^{-\gamma_{\vec{p}}}\gamma_{\vec{p}}^2+16e^{-\frac{1}{4}\gamma_{\vec{p}}}\gamma_{\vec{p}}^2+8e^{-\frac{1}{2}\gamma_{\vec{p}}}\gamma_{\vec{p}}^2+4e^{-4\gamma_{\vec{p}}}-8e^{-\gamma_{\vec{p}}}+2e^{-2\,\gamma_{\vec{p}}}+\\
&+2e^{-\frac{1}{4}\gamma_{\vec{p}}}\gamma_{\vec{p}}^3\bigg).
\end{split}
\end{equation}
The expression above is not very illuminating in this form but it again shows that the energy of ``one-particle" states obeys a deformed dispersion relation. If we expand for $\gamma_{\vec{p}}=\omega_{\vec{p}}\ell\ll1$ we obtain the leading order polymer dispersion relation
\begin{equation}
E_{\vec{p}}^2 \simeq m^2+ |\vec{p}|^2+\frac{3\ell |\vec{p}|^3}{4}\,,
\end{equation}
which shows that the leading order correction to the one-particle dispersion relation in comparison to the vacuum result (\ref{disp_rel}) is still cubic in the modulus of spatial momentum but with with a different numerical coefficient. The appearance of these deformed dispersion relations suggests a violation of Lorentz symmetry due to the presence of the polymer scale as indeed it was pointed out in \cite{pqft1,pqft2}. It is easy to see that for $\gamma_{\vec{k}}\ll1$ all the ordinary QFT expectation values are recovered and, following the discussion in \cite{pqft1}, this justifies the definition of the vacuum state and the Fock operators adopted here. In the next sections we will try to capture the physical meaning of polymer quantization by analyzing two simple procedures of localization  in a field theoretic setting. Our first step in the next Section will be to review two possible approaches to localization in ordinary quantum field theory which we will then generalize to the polymer case.

\section{Localization in quantum field theory}
In ordinary QFT particles are described by asymptotic Fock space states in a scattering process. Such states are irreducible representations of the Poincar\'e group and are labelled by their eigenvalues with respect to a maximal set of commuting generators of Poincar\'e transformations. The usual choice for such set is the spatial translation generators so that the quantum representative of a free relativistic particle with linear momentum $\vec{k}$ and energy $\omega(\vec{k})$ will be the ``ket" $|\vec{k}\rangle$.  This particle picture of course knows nothing about the region where the particle is localized and as a matter of fact Fock space states correspond to {\it global} one-particle states since they are {\it not} eigenstates of {\it local} observables namely operators defined in a finite region of observation.\\
In order to operationally address the issue of localization for a particle or a measurement in PQFT we resort here to two strategies.  The first one will be to simply impose boundary conditions for field operators and their eigenstates so that they have support on a given ``local" volume $L$ representing the physical size of the measuring apparatus.  
A different approach to localization will be to introduce {\it localized} one-particle states by considering appropriate superpositions of one particle states weighted by a localizing {\it profile} function without affecting the boundary conditions of the observables.

\subsection{Local vs. global: boundary conditions}
We start by outlining the first strategy along the lines of \cite{Colosi:2004vw} and working out the tools which we will need for our extension to PQFT.  The basic idea will be to tune the boundary conditions of the fields and/or the integration region of the Hamiltonian density using the \textit{global} regularizing volume $V$ or the \textit{local} volume $L$. Then we explicitly evaluate the expectation values of local and global Hamiltonians in both local and global states.\\
To keep our discussion simpler we consider a scalar field in two space-time dimensions, defined in box volume of size $V$, with boundary conditions $\phi(-V/2)=\phi(V/2)$. The Hamiltonian is
\begin{equation}\label{global_ham}
H=\frac{1}{2}\int_{-V/2}^{V/2} dx \bigg[\pi^2+\bigg(\partial_x \phi\bigg)^2+m^2\phi^2\bigg].
\end{equation}
where $\pi(x,t)$ is the momentum conjugate to $\phi(x,t)$. A orthogonal set of mode functions is given by
\begin{equation}
u_k(x,t)\propto e^{+\imath kx-\imath \omega_k t},
\end{equation}
where $k=\frac{2\pi n}{V}$, $n\in\mathbb{Z}$ label the discrete modes of the system and $\omega_k^2=\bigg(\frac{2\pi n}{V}\bigg)^2+m^2$ is the energy of the individual mode.  Using the standard Klein-Gordon inner product we can choose a normalization constant
\begin{equation}
\bigg(u_k(x,t),u^{\dagger}_k(x,t)\bigg)=-\imath\int_{-V/2}^{V/2} dx \big( u_k \partial_t u^{\dagger}_k -\big( \partial_t u_k\big)u_k^{\dagger}\big)=2\omega_k V,
\end{equation}
\begin{equation}\label{modes}
\Rightarrow u_k(x,t)=\frac{1}{\sqrt{2\omega_k V}}e^{+\imath kx-\imath \omega_k t}.
\end{equation}
At a quantum level we can perform a standard expansion of the field and the momentum in terms of the modes above using creation and annihilation operators
\begin{equation}\label{field}
\phi(x,t)=\sum_k \bigg[a_k u_k(x,t)+a^{\dagger}_k u^{\dagger}(x,t)\bigg],\,\,\,\,\,\,
\pi(x,t)=\partial_0 \phi(x,t)=\sum_k (-\imath \omega_k)\bigg[a_k u_k(x,t)-a^{\dagger}_k u^{\dagger}(x,t)\bigg],
\end{equation}
obeying the usual canonical commutation relations which lead to the well known expression for the Hamiltonian 
\begin{equation}\label{global_ham_fock}
H=\sum_k \omega_k a^{\dagger}_k a_k.
\end{equation}
Let us now consider a detector located in a finite region $L<V$. The energy measured by such device will be given by the expectation value of the Hamiltonian density integrated over the region $L$
\begin{equation}\label{loc_ham_int}
H_L=\frac{1}{2}\int_{-L/2}^{L/2} dx \bigg[\pi^2+\bigg(\partial_x \phi\bigg)^2+m^2\phi^2\bigg].
\end{equation}
The operator $H_L$ can be written in terms of the ``global" creation and annihilation operators $a_k$ and $a_k^{\dagger}$ simply substituting the expression of the field and the conjugate momentum into (\ref{loc_ham_int}) 
\begin{equation}\label{impl_loc_ham}
H_L=\sum_{k,k'}\bigg(A_{k,k'}a_k^{\dagger}a_{k'}^{\dagger}+B_{k,k'}a_k a_{k'}+C_{k,k'}a_k^{\dagger}a_{k'}+D_{k,k'}a_k a_{k'}^{\dagger}\bigg),
\end{equation}
where the matrices $A,B,C,D$ are computed from the ``overlap" integrals
\begin{equation}
U_{k,k'}=\int_{-L/2}^{L/2} dx\, u_k(x,t) u_{k'}(x,t)\, , 
\end{equation}
which reduce to discrete deltas if the region spanned by the integral is equal to the support of the modes. A straightforward but tedious calculation gives us the explicit form of $H_L$ in terms of the global creation and annihilation operators
\begin{equation}\label{loc_ham}
\begin{split}
H_L=&\frac{1}{2}\sum_{k,k'}\bigg\{ \bigg( a^{\dagger}_k a^{\dagger}_{k'}e^{\imath(\omega_k+\omega_{k'})t}+ 
a_k a_{k'}e^{-\imath(\omega_k+\omega_{k'})t}\bigg)\bigg[ \frac{(-\omega_k \omega_{k'}-kk'+m^2)}{2\sqrt{\omega_k\omega_{k'}}}\bigg]\frac{\sin(\frac{(m+n)L\pi}{V})}{(m+n)\pi}+\\
&+\bigg( a_k a^{\dagger}_{k'}e^{-\imath(\omega_k-\omega_{k'})t}+a^{\dagger}_k
a_{k'}e^{\imath(\omega_k-\omega_{k'})t}\bigg)\bigg[ \frac{(\omega_k \omega_{k'}+kk'+m^2)}{2\sqrt{\omega_k\omega_{k'}}}\bigg]\frac{\sin(\frac{(m-n)L\pi}{V})}{(m-n)\pi}\bigg\}.
\end{split}
\end{equation}
It is evident that the local Hamiltonian $H_L$ does not commute with the ``global" number operator thus it contains terms that map out of the one particle subspace.  Thus global one particle Fock states cannot be eigenstates of this operator. To find the eigenestates of $H_L$ we can write a set of modes restricted to the subspace $L$. We could use reflecting boundary conditions $\phi(-L/2)=\phi(L/2)=0$ (a field constrained in a box). This choice would force us to use $\sin$ ($\cos$) functions as modes of the system. We use instead, without loss of generality, plane-wave modes in order to simplify calculations.  These local modes are given by
\begin{equation}\label{local_modes}
u_k^L(x,t)=\frac{1}{\sqrt{2\omega_k^L L}}e^{+\imath kx-\imath \omega_k^L t},
\end{equation}
with $k=\frac{2\pi n}{L}$ and their dispersion relation is given by
\begin{equation}\label{local_energy}
\omega_k^{L 2}=(\frac{2\pi n}{L})^2+m^2.
\end{equation}
We call {\it local} one-particle states the excitations of the modes which have support on the region $L<V$. A {\it localized} field is given by an expansion in terms of local annihilation and creation operators and the modes (\ref{local_modes})
and correspondingly the local Hamiltonian $H_L$ will read
\begin{equation}
H_L=\sum_k \omega_k^L a^{L\dagger}_k a_k^L.
\end{equation} 
It is instructive to evaluate the expectation value of (\ref{loc_ham}) on generic global one particle states $|q\rangle$ and $|p\rangle$
\begin{equation}
\langle q|H_L|p\rangle=\frac{1}{2}\frac{L}{V}\sum_k \omega_k+\bigg[\frac{\omega_q\omega_p+qp+m^2}{2\sqrt{\omega_p\omega_q}}\bigg]\frac{\sin\big(\frac{(m-n)\pi L}{V}\big)}{(m-n)\pi}\cos\left((\omega_p-\omega_q)t\right).
\end{equation}
As anticipated above $H_L$ is non-diagonal in the basis of ``global" one-particle states, nevertheless we can evaluate the expectation value above for $q\rightarrow p$, i.e. for a given one particle momentum
\begin{equation}\label{exp_loc_ham}
\langle p|H_L|p\rangle=\frac{L}{V}\omega_p +\frac{1}{2}\frac{L}{V}\sum_k \omega_k.
\end{equation}
Quite sensibly we find that the expectation value of (\ref{loc_ham_int}) in a global one particle state is the energy density $\frac{\omega_p}{V}$ times the local volume $L$. This result is in agreement with the interpretation that the expectation value of the operator (\ref{loc_ham_int}) represents the portion of energy contained in the sub-region $L$. The first term is the one particle contribution; the second term is a summation over vacuum contributions, one for each oscillator.  In the limit of the localization volume going to zero the local energy vanishes and in the limit $L\rightarrow V$ the local energy reduces to the global particle energy $\omega_p$ as one would expect.\\ 
At this point one could suggest that according to (\ref{local_energy}) the energy of a localized state, as intended in this paragraph, is divergent for $L\rightarrow 0$. But that relation has nothing to do with a localization procedure. It is simply telling us that if we take a field defined in a smaller box, we have to spend more energy in order to cause a transition between two different states $n$ and $n'>n$.
Let us notice in closing that if we evaluate the expectation value of the global Hamiltonian $H$ (which is the Hamiltonian density integrated over a global region $V$) on a local particle state $a^{L\dagger}_k|0\rangle=|k\rangle_{L}$ we simply obtain the energy of the excitations contained in the subregion $L$ namely $H_V|p\rangle_L = \omega_p^{L} |p\rangle_L$\footnote{In order to get this result one has to impose that $u_k^L (x,t)=0$ outside the local region $L$.}.

\subsection{Localized states}
A different picture of localization for a quantum relativistic particle makes use of appropriate superpositions of one particle states.  In particular we can consider the ``wave-function" ket  obtained by acting with the field operator on vacuum $|\phi\rangle=\phi(\vec{x},t)|0\rangle$ and define a localized state
\begin{equation}
|\Phi(f)\rangle_{\mathrm{loc}}=\int d^3 x f(\vec{x}-\vec{x}_0,\sigma)\phi(\vec{x})|0\rangle,
\end{equation}
where $f(\vec{x}-\vec{x}_0,\sigma)$ is a {\it window function} with support on all space, i.e. on the global region $V$, and centered at the point $\vec{x}_0$.  We can interpret the introduction of this function as a way of separating the volume of the system into two subspaces, an observable region of ``size" $\sigma$ and an unobservable one, without affecting the boundary conditions of the field.
As before, we are interested in evaluating the energy in the subsystem represented by the localized state.  For simplicity we look at a state centered at the origin and consider its expression in momentum space
\begin{equation}\label{smoothed_state}
|\Phi(f)\rangle_{\mathrm{loc}}=\int d^3 x f(\vec{x},\sigma)\phi(\vec{x})|0\rangle=\\
\int \frac{d^3 p}{\sqrt{(2\pi)^3 2 \omega_p}} f(\vec{p},\sigma)e^{\imath p_0 t}|p\rangle,
\end{equation}
from which one straightforwardly obtains
\begin{equation}\label{smoothed_ham}
\phantom{}_{\mathrm{loc}}\langle\Phi(f)|H|\Phi(f)\rangle_{\mathrm{loc}}=\int d^3 p f(\vec{p},\sigma) |p| f(-\vec{p},\sigma)=\int d^3 x f(\vec{x},w) \sqrt{\vec{\nabla}^2}f(-\vec{x},\sigma).
\end{equation}
We specialize to a Gaussian window function $f(\vec{x},\sigma)\propto e^{-\frac{(\vec{x})^2}{2\sigma^2}}$ and a massless free scalar field with $\omega_{p}=|p|^2$ in two space-time dimensions. Taking into account the normalization given by $\phantom{}_{\mathrm{loc}}\langle \Phi|\Phi\rangle_{\mathrm{loc}}$,  one finds 
\begin{equation}\label{energy_gauss}
\frac{\phantom{}_{\mathrm{loc}}\langle \Phi|H|\Phi\rangle_{\mathrm{loc}}}{\phantom{}_{\mathrm{loc}}\langle \Phi|\Phi\rangle_{\mathrm{loc}}}=\left(\frac{\sigma}{\sqrt{\pi}}\right)\int dp\,e^{-p^2 \sigma^2}|p|=\frac{1}{\sqrt{\pi} \sigma}.
\end{equation}
If we interpret $\sigma$ as the uncertainty on the location of the particle then (\ref{energy_gauss}) can be interpreted in terms of the familiar relation $E\sim1/\delta x$ which is nothing but the expression of the ``relativistic uncertainty relation" for a gaussian wavepacket \cite{landau_p}.

\section{Polymer fields and localization}
In the following sections we explore in detail the extension of the two procedures of localization discussed above to polymer quantum fields. As we will see while conceptually straightforward this extension requires complex manipulation of expectation values of polymer operators. The final results are however quite intuitive and suggest that polymer effects 
do not alter the notion of localized states and observables at the one-particle level.

\subsection{Local boundary conditions in Polymer QFT}
We start by considering the measurement of an observable restricted to a local volume. According to our previous discussion we focus on the local Hamiltonian (\ref{loc_ham}) and consider its polymer counterpart. In the approach we are considering in this work polymer quantization affects the creation and annihiliation operators while manipulations with field modes can be carried out in the usual fashion. The polymer local Hamiltonian can be thus obtained following the same steps that led  to (\ref{loc_ham}) but replacing ordinary creation and annihilation operators with polymer ones 
\begin{equation}\label{loc_ham_pol}
\begin{split}
H_L=&\frac{1}{2}\sum_{k,k'}\bigg\{ \bigg( A^{\dagger}_k A^{\dagger}_{k'}e^{\imath(\omega_k+\omega_{k'})t}+ 
A_k A_{k'}e^{-\imath(\omega_k+\omega_{k'})t}\bigg)\bigg[ \frac{(-\omega_k \omega_{k'}-kk'+m^2)}{2\sqrt{\omega_k\omega_{k'}}}\bigg]\frac{\sin(\frac{(m+n)L\pi}{V})}{(m+n)\pi}+\\
&+\bigg( A_k A^{\dagger}_{k'}e^{-\imath(\omega_k-\omega_{k'})t}+A^{\dagger}_k
A_{k'}e^{\imath(\omega_k-\omega_{k'})t}\bigg)\bigg[ \frac{(\omega_k \omega_{k'}+kk'+m^2)}{2\sqrt{\omega_k\omega_{k'}}}\bigg]\frac{\sin(\frac{(m-n)L\pi}{V})}{(m-n)\pi}\bigg\}.
\end{split}
\end{equation}
We are interested in the expectation value of the ``local" Hamiltonian (\ref{loc_ham_pol}) on the ``global" polymer one-particle state $|p\rangle$.
One can show that $\langle p|H_L(A_k,A_k^{\dagger},A_{k'},A_{k'}^{\dagger})|p\rangle\neq0$ only for $k=k'$, therefore we have that  $\langle n_k|(A^{\dagger}_k A^{\dagger}_k)^N|n_k\rangle=\langle n_k|(A_k A_k)^N|n_k\rangle=0$ for $N\neq 0$, where $|n_{k}\rangle\propto\big(A^{\dagger}_{k}\big)^n |0_{k}\rangle$. Thus the only non vanishing terms are $\langle p|A_k A^{\dagger}_{k'}|p\rangle$ and $\langle p|A^{\dagger}_k A_{k'}|p\rangle$. After some tedious but straightforward calculations we obtain
\begin{equation}
\begin{split}
&\langle p|H_L|p\rangle=\frac{1}{2}\bigg\{\frac{L}{V}\bigg[\omega_p \frac{\langle 0|A_p(A_p A^{\dagger}_p)A^{\dagger}_p|0\rangle+\langle 0|A_p(A^{\dagger}_p A_p)A^{\dagger}_p|0\rangle}{\langle p|p\rangle}+\\
&+\sum_{k\neq p}\omega_k\bigg(\langle 0|A_k A^{\dagger}_k|0\rangle+\langle 0_k|A^{\dagger}_k A_k|0_k\rangle\bigg)\bigg]\bigg\}.
\end{split}
\end{equation}
Working out the explicit form of the expectation values and subtracting the vacuum energy contribution we have our final result
\begin{equation}\label{loc_en}
\begin{split}
&\langle p|H_L|p\rangle=\frac{L}{V}\bigg[\frac{1}{\langle p|p\rangle}\langle 0|A_p[H(A_k,A^{\dagger}_k)]A^{\dagger}_p|0\rangle\bigg]=\\
&=\frac{L}{V}\bigg[\frac{\omega_p}{16\gamma_p \bigg(-e^{-\gamma_p}+1+2e^{-\frac{1}{4}\gamma_p}\gamma_p+\gamma_p \bigg) }\bigg(5+12\gamma_p-6e^{-\frac{9}{4}\gamma_p}\gamma_p+\\
&+10e^{-\frac{5}{4}\gamma_p}\gamma_p-4e^{-\frac{5}{4}\gamma_p}\gamma_p+4e^{-\gamma_p}\gamma_p^2+16e^{-\frac{1}{4}\gamma_p}\gamma_p^2+8e^{-\frac{1}{2}\gamma_p}\gamma_p^2+4e^{-4\gamma_p}+\\
&-8e^{-\gamma_p}+2e^{-2\,\gamma_p}+2e^{-\frac{1}{4}\gamma_p}\gamma_p^3\bigg)\bigg].
\end{split}
\end{equation}
While the resulting expression looks quite complicated it has a rather simple physical interpretation. Indeed considering (\ref{oneparticle_energy_polymer}), we can rewrite this result as
\begin{equation}
\langle p|H_L|p\rangle=\frac{L}{V}\langle p|H|p\rangle\,,
\end{equation}
which tells us that the expectation value of the localized Polymer Hamiltonian is given by the polymer {\it energy density} $\langle p|H|p\rangle/V$ multiplied by the local volume $L$. 
Therefore we found that, {\it mutatis mutandis}, we encounter the same situation as in ordinary QFT: the expectation value of a local polymer Hamiltonian in a global polymer one particle state can be seen as the portion of energy contained in the local region.

\subsection{Polymer smeared states}
\begin{figure}[htbp]
\begin{center}
\includegraphics[width=9cm, height= 6cm]{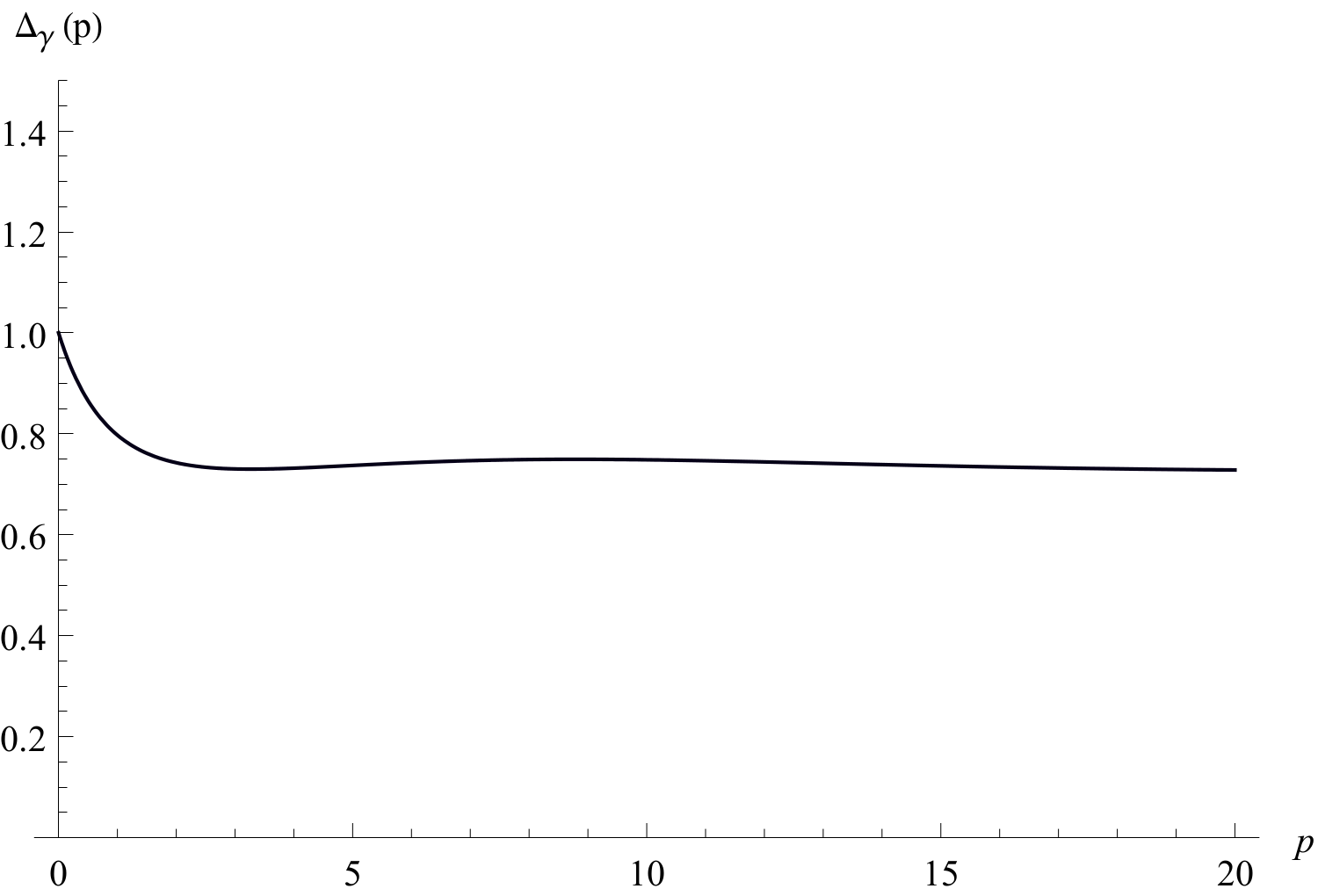}
\caption{Polymer correction to the ordinary one particle dispersion relation.}
\label{pol_delta}
\end{center}
\end{figure}

\begin{figure}[htbp]
\begin{center}
\includegraphics[width=9cm, height= 6cm]{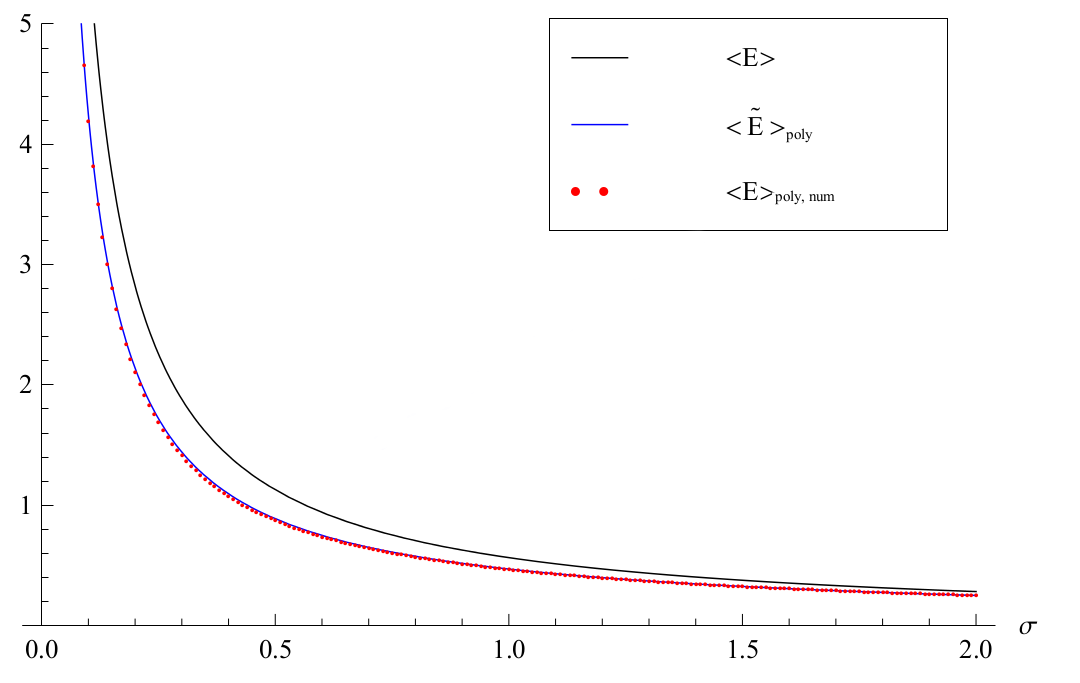}
\caption{Expectation value of the Energy for a smoothed state in the ordinary case (continuous black line) and in the polymer case (continuous blue line and dotted line).}
\label{3x}
\end{center}
\end{figure}

We turn now to the case of polymer localized states constructed from a superposition of polymer one-particle states weighted by a suitable profile function with support on the localization region. Also in this case our goal is to establish how polymer quantization affects the properties of a basic observable like the Hamiltonian when the size of the localization region reaches polymer scales. We know from (\ref{smoothed_state}) that a smoothed state can be written in momentum space as
\begin{equation}
|f\rangle=\int dk f(k) |k\rangle,
\end{equation}
where $f(k)$ is the Fourier transform of the profile function $f(x)$. In analogy with the undeformed case we consider a gaussian profile function and the associated polymer {\it localized} one-particle state can be written as
\begin{equation}\label{pol_gauss}
|\phi\rangle=\sum_k G(k,\sigma) \frac{A^{\dagger}_k}{\sqrt{\langle p|p\rangle}} |0\rangle,
\end{equation}
where $G(k,\sigma)=e^{-\frac{p^2\sigma^2}{2}}$ is the Fourier transform of a Gaussian in coordinate space. Our goal is the evaluate the expectation value of the polymer Hamiltonian in such state
\begin{equation}
\langle \phi|H|\phi\rangle=\sum_p\sum_{p'}\langle p|G(p,\sigma)\bigg[\sum_{k}\omega_k \bigg(A_{k}^{\dagger}A_{k}+\frac{1}{2}\big[A_{k},A_{k}^{\dagger}\big] \bigg)\bigg]G(p',\sigma')|p'\rangle.
\end{equation}
As in the calculations above we notice that only the $p=p'=k$ and the $p=p'\neq k$ terms contribute to the sum. Using integrals instead of summations we obtain\footnote{We do not care about numerical factors because we are going to normalize the result.}
\begin{equation}
\langle\phi|H|\phi\rangle\simeq\int dp\,G^2(p,\sigma)\omega_p\langle p|\big(A_{p}^{\dagger}A_{p}+\frac{1}{2}\big[A_{p},A_{p}^{\dagger}\big] \big)|p\rangle+\int dp\,G^2(p,\sigma)\int dk \frac{\omega_k}{2}\big(\langle 0|A^{\dagger}_k A_k|0\rangle+\langle 0|A_k A^{\dagger}_k|0\rangle\big),
\end{equation}
where the second term is the usual vacuum contribution. We can use the results of the previous section to write the first term. If we denote with $\Delta_\gamma(p)$ (see Figure \ref{pol_delta}) the polymer correction to the one particle energy, i.e. $\langle p|H|p\rangle=\omega_p\Delta_\gamma (p)$, the expectation value of the polymer Hamiltonian on the localized state can be written as 
\begin{equation}\label{exact_int}
\langle\phi|H|\phi\rangle\simeq\int dp\,G^2(p,\sigma)\omega_p\Delta_\gamma(p),
\end{equation}
where
\begin{equation}\label{exact_pol_corr}
\begin{split}
&\Delta_\gamma (p)=\frac{1}{16\gamma_p \bigg(-e^{-\gamma_p}+1+2e^{-\frac{1}{4}\gamma_p}\gamma_p+\gamma_p \bigg) }\bigg(5+12\gamma_p-6e^{-\frac{9}{4}\gamma_p}\gamma_p+10e^{-\frac{5}{4}\gamma_p}\gamma_p+\\
&-4e^{-\frac{5}{4}\gamma_p}\gamma_p+4e^{-\gamma_p}\gamma_p^2+16e^{-\frac{1}{4}\gamma_p}\gamma_p^2+8e^{-\frac{1}{2}\gamma_p}\gamma_p^2+4e^{-4\gamma_p}-8e^{-\gamma_p}+2e^{-2\,\gamma_p}+2e^{-\frac{1}{4}\gamma_p}\gamma_p^3\bigg).
\end{split}
\end{equation}
The integral in (\ref{exact_int}) can not be analytically solved and thus we resorted to numerical techniques. We evaluated
\begin{equation}
\langle E\rangle_{poly}=\frac{\langle\phi|H|\phi\rangle}{\langle\phi|\phi\rangle}=\left(\frac{\sigma}{\sqrt{\pi}}\right)\int dp\,G^2(p,\sigma)\omega_p\Delta_\gamma(p)\,,
\end{equation}
setting $\ell=1$ and varying the width of the gaussian $\sigma$ between $0.01$ and $10$ in steps of $0.01$. 
\begin{figure}[htbp]
\begin{center}
\includegraphics[width=9cm, height= 6cm]{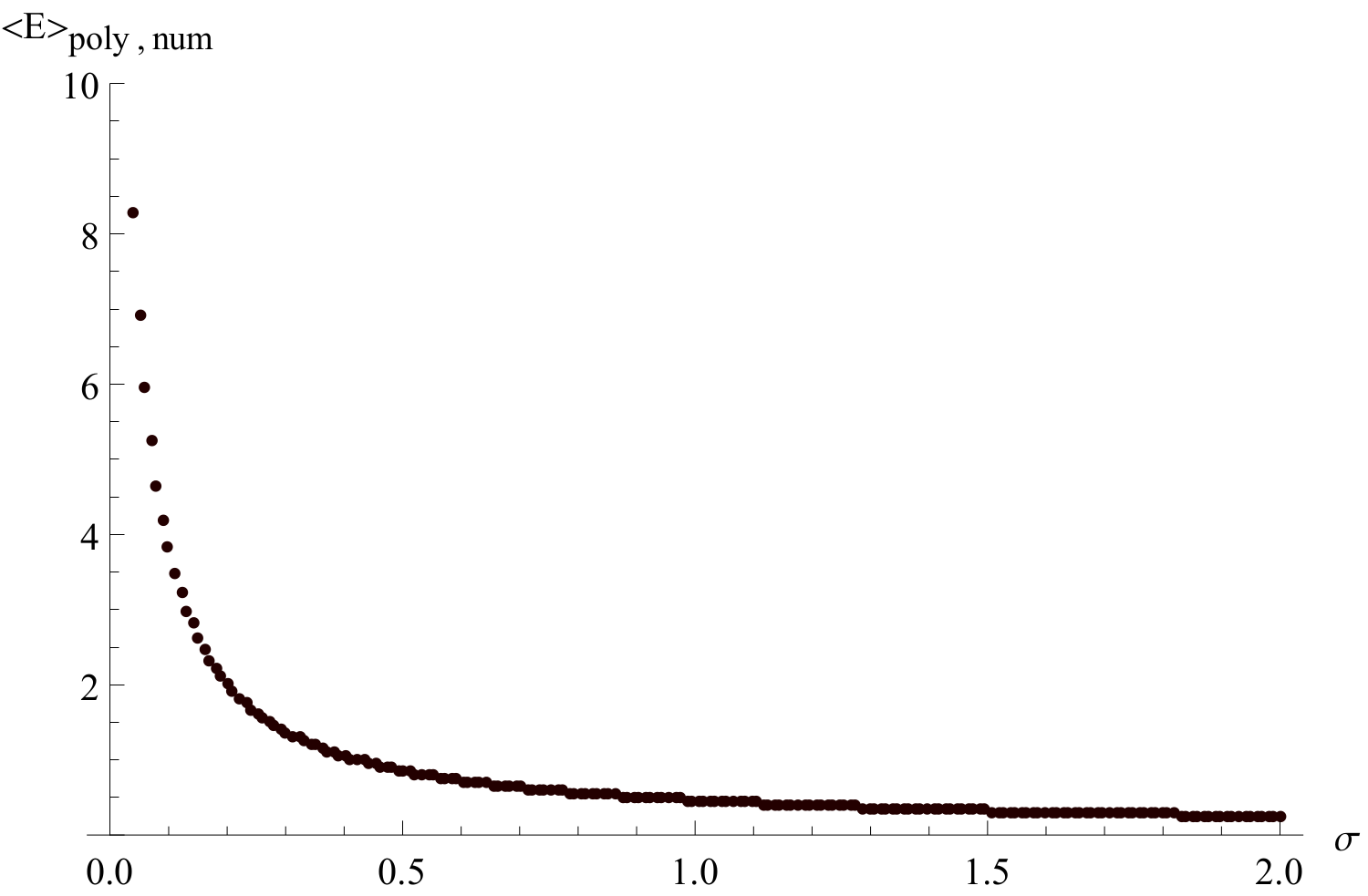}
\caption{Numerical evaluation of $\langle E\rangle_{poly}$ for $\ell=1$ and $0.01<\sigma<2$, $\Delta\sigma=0.01$.}
\label{pol_num}
\end{center}
\end{figure}
Notice that the range that we have chosen for $\sigma$ goes well below the characteristic scale $\ell$. As one can see from Fig. \ref{pol_num} below $<E>_{poly}$ behaves qualitatively as in the ordinary case.\\ We can proceed further in the analysis of $<E>_{poly}$ by noticing from figure \ref{pol_delta} that $\Delta_\gamma(p)$ has a quite simple profile that can be approximated by a decreasing exponential for $0<p\lesssim 1 $ and by a constant for $p>1$. Therefore we can replace $\Delta_\gamma(p)$ with a test function $\tilde{\Delta}_\gamma(p)$ that mimics its behaviour. For example we can take
\begin{equation}
\tilde{\Delta}_\gamma(p)=\frac{3}{4}+\frac{1}{4}e^{-\frac{3}{2}|p|\ell}.
\end{equation}
$\tilde{\Delta}_\gamma(p)$ has the same expansion of $\Delta_\gamma(p)$ for $\gamma_p\ll1$ except for numerical coefficients beyond first order in $\gamma_p$
\begin{equation}
\begin{split}
&\Delta_\gamma(p)\simeq 1-\frac{3\gamma_p}{8}+\frac{53}{192}\gamma_p^2+\mathcal{O}(\gamma_p^3),\\
&\tilde{\Delta}_\gamma(p)\simeq 1-\frac{3\gamma_p}{8}+\frac{9}{32}\gamma_p^2+\mathcal{O}(\gamma_p^3),
\end{split}
\end{equation}
and the same limit for $\gamma_p\rightarrow\infty$
\begin{equation}
\begin{split}
&\Delta_\gamma(p)\underset{\gamma_p\rightarrow\infty}{\longrightarrow}\frac{3}{4},\\
&\tilde{\Delta}_\gamma(p)\underset{\gamma_p\rightarrow\infty}{\longrightarrow}\frac{3}{4}\,,
\end{split}
\end{equation}
which allows us to explore in more detail the UV and IR limiting behaviours of $<E>_{poly}$.  It is possible control the error we make replacing $\Delta_\gamma(p)$ with $\tilde{\Delta}_\gamma(p)$ simply considering the quantity
\begin{equation}
Err(\sigma,\ell)=1-\frac{\int_{-\infty}^{+\infty} dp |p|\tilde{\Delta}_\gamma(p)e^{-p^2\sigma^2}}{\int_{-\infty}^{+\infty} dp |p|\Delta_\gamma(p)e^{-p^2\sigma^2}}.
\end{equation}
We calculated it numerically fixing $\ell=1$ for $0.01<\sigma<10$ with steps of $0.01$ and $10^{-4}<\sigma<0.1$ with steps of $5\cdot 10^{-4}$.
Within the interval $10^{-4}<\sigma<10$ we have that $Err(\sigma,1)<0.023$. That means that the gaussian integral with the polymer dispersion relation is well approximated by the one with the test function and the percentage error is less than $0.023\%$ in the interval considered. Thus we are reassured that $\tilde{\Delta}_\gamma (p)$ is a good candidate test function and we can proceed to analytically calculate the following integral
\begin{equation}
<\tilde{E}>_{poly}=\left(\frac{\sigma}{\sqrt{\pi}}\right)\int_{-\infty}^{+\infty} dp\,e^{-p^2\sigma^2}|p|\left(\frac{3}{4}+\frac{1}{4}e^{-\frac{3}{2}|p|\ell}\right),
\end{equation}
where $\omega_p=|p|$.
This is simply a Gaussian integral on the positive axis, with a first moment and a linear term which evaluated gives
\begin{equation}\label{pol_smoothed}
<\tilde{E}>_{poly}=\frac{1}{\sqrt{\pi}\sigma}\left(1-\frac{3\sqrt{\pi}\ell\,e^{\frac{9\ell^2}{16\sigma^2}}\,Erfc\left(\frac{3\ell}{4\sigma}\right)}{16\sigma}\right),
\end{equation}
where $Efrc\left(\frac{3\ell}{4\sigma}\right)$ is the complementary error function and it is defined as
\begin{equation}
Erfc(x)=\frac{2}{\sqrt{\pi}}\int_{x}^{\infty}\,e^{-t^2}\,dt.
\end{equation}
It has the special values $Erfc(-\infty)=2,\;Erfc(0)=1,\;Erfc(+\infty)=0$.\\
Let us explore the UV regime of the theory, at scales much smaller than the polymer scale, by taking the ``extreme localization" limit $\sigma\rightarrow 0$. We evaluate the power series for $\sigma\ll\ell$ of (\ref{pol_smoothed})
\begin{equation}\label{E_poly}
<\tilde{E}>_{poly}=\frac{3}{4\sqrt{\pi}\sigma}+\mathcal{O}(\sigma)\, ,\,\sigma\ll\ell.
\end{equation} 
We can see how the average energy diverges with $\sim\frac{1}{\sigma}$ as the localization parameter goes to zero, in complete analogy with the ordinary field theory case. This is because the term $\left(3\ell\,e^{\frac{9\ell^2}{16\sigma^2}}\,Erfc\left(\frac{3\ell}{4\sigma}\right)\right)/16\sigma^2$ is such that for $\sigma\ll1$ it does not compensate the $\frac{1}{\sqrt{\pi}\sigma}$ term and it gives a contribution only for $\sigma\sim\ell$.\\
To probe the IR behaviour we expand (\ref{pol_smoothed}) for $\ell\ll\sigma$ and derive the the first order correction in $\ell$
\begin{equation}\label{E_poly_pert}
<\tilde{E}>_{poly}=\frac{1}{\sqrt{\pi } \sigma }-\frac{\ell}{4 \sigma ^2}+\mathcal{O}(\ell^2)\,,\,\ell\ll\sigma\, ,
\end{equation}
which, as one would expect also from dimensional analysis, goes as $1/\sigma^2$ and thus is vanishingly small, as expected. In Fig. \ref{3x} we comparatively plot the expectation value of the energy of a localized state for polymer and ordinary field theory both for our numerical results using $\Delta_\gamma(p)$ and for the analytical results using the test function $\tilde{\Delta}_\gamma(p)$. We see that polymer effects are relevant only in the ``polymer" region $\sigma\sim \ell$ and do not affect the behaviour of the energy of the localized state significantly at other scales. Finally we notice that the polymer scale {\it does not} act as an effective minimal length scale since localized states at sub-polymer scales are still allowed if one allows unbounded growth of the energy of the state.

\section{Polymer diffusion and spectral dimension}
\begin{figure}[htbp]
\begin{center}
\includegraphics[width=9cm, height= 6cm]{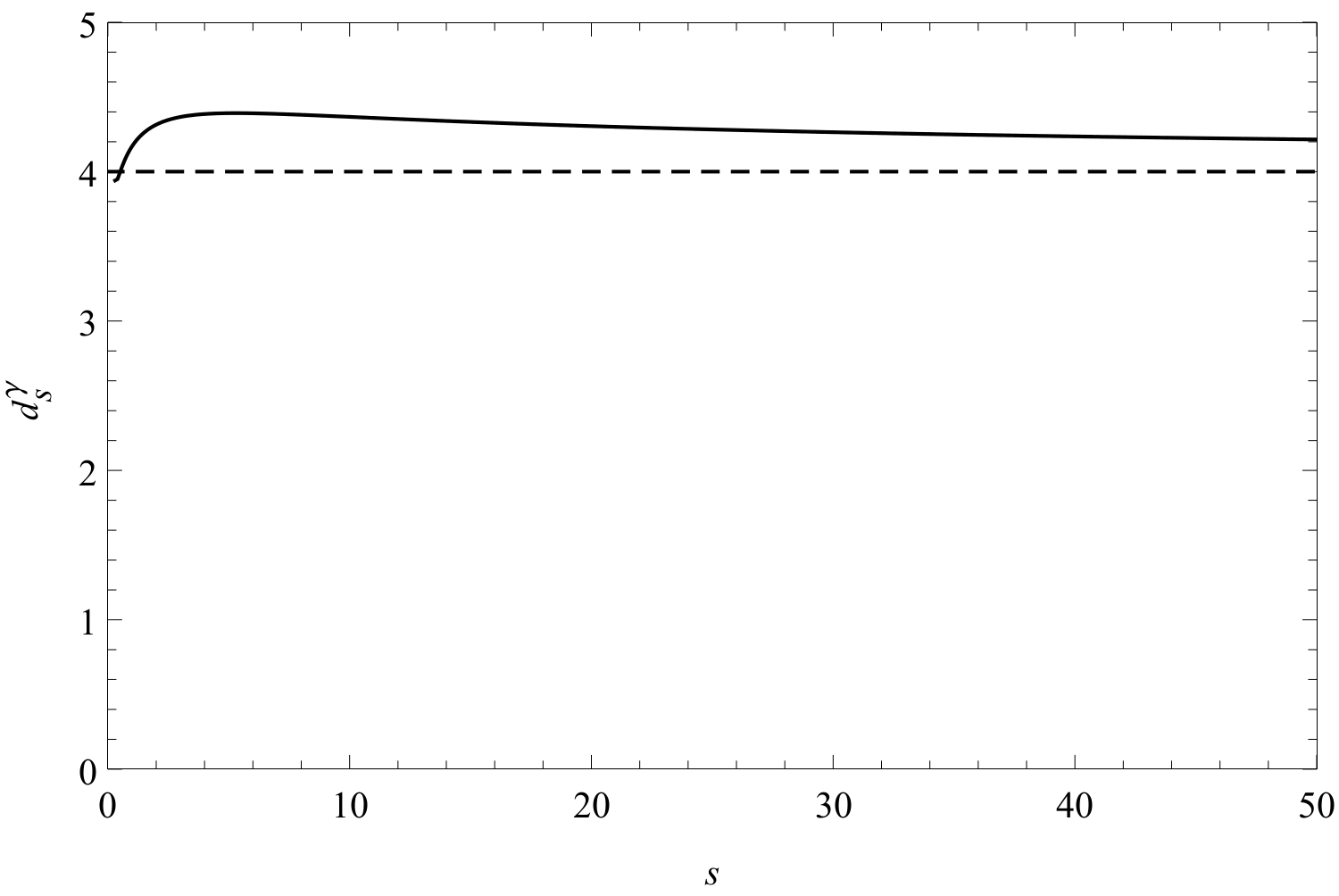}
\caption{Spectral dimension in units $\ell=1$ (the continuous line, the dashed line is for ordinary Euclidean space)}
\label{pol_spectr}
\end{center}
\end{figure}

Having shown that polymer corrections affect only mildly the two localization procedures we considered so far we turn in this section to a subtler tool for exploring the effective space-time picture that emerges in our polymer setting. This consists in the use of a ``fictitious" diffusion process governed by an effective deformed Laplacian carrying corrections due to quantum gravity effects. The associated notion of spectral dimension has proved valuable in gaining insight on the small scale structure of space-time in various quantum gravity scenarios where no other tools are available or where a notion of space-time as a continuum manifold is lacking \cite{Ambjorn:2005db,Lauscher:2005qz,Horava:2009if,Benedetti:2008gu,Modesto:2008jz,Carlip:2009kf,Calcagni:2010pa,Sotiriou:2011mu,Alesci:2011cg,Amelino-Camelia:2013cfa,Arzano:2014jfa}.\\
In general if we start from some non-conventional (non-local, Lorentz breaking, non-commutative etc.) field theory, where a (generalized) Laplacian operator $\Delta_{\vec{x}}$ is available, we can consider the {\it heat equation} 
\begin{equation}
\partial_s K(\vec{x},\vec{x}';s)+\Delta_{\vec{x}} K(\vec{x},\vec{x}';s)=0.
\end{equation}
Here $s$ is an ``auxiliary" diffusion time and $K(\vec{x},\vec{y};s)$ is the heat kernel which solves the equation above. In most cases the starting theory possesses a deformed d'Alembertian operator which is Wick rotated to the deformed Laplacian $\Delta_{\vec{x}}$. For flat Minkowski space-time we have, for example
\begin{equation}
K(\vec{x},\vec{x}';s)=\int \frac{d^n p}{(2\pi)^n}e^{-\omega_{\vec{p}}^2\,s}e^{\imath \vec{p}\cdot(\vec{x}-\vec{x}')},
\end{equation}
where $\omega_{\vec{p}}$ is the ``Euclidean" dispersion relation and $n$ is the number of spatial dimensions\footnote{We are only considering ghost-free models where the generalized (Euclidean) differential operator governing the equations of motion depends trivially on time, as $D=-\partial_t ^2+f(-\Delta_{\vec{x}})$. In this case $d_s$ refers to the spatial part of the spectral dimension.}. We obtain the trace of the heat kernel taking $\vec{x}=\vec{x}'$. It reads as
\begin{equation}\label{trace}
Tr\,K(\vec{x},\vec{x}';s)=P(s)=\int \frac{d^n p}{(2\pi)^n}e^{-\omega_{\vec{p}}^2 s}\,,
\end{equation}
and can be interpreted as the {\it return probability} for the diffusion process (see e.g. \cite{Calcagni:2013vsa}). To this quantity we can associate a notion of \textit{spectral dimension} defined by 
\begin{equation}\label{spectr}
d_s=-2\frac{\partial \log P(s) }{\partial\log s}.
\end{equation}
In ordinary Euclidean (Wick-rotated Minkowski) space it can be easily verified that the spectral dimension does not actually depend on the diffusion parameter $s$ i.e. is {\it scale independent} and its constant value $n$ coincides with the Hausdorff dimension i.e. the scaling of the volume of a $n$-ball with given radius. In more non-trivial settings $d_s$ exhibits a ``running" with the scale which in the IR ($s\rightarrow\infty$) is due to curvature effects while in the UV ($s\rightarrow 0$) is determined by non-trivial small-scale structures of space-time \cite{Sotiriou:2011aa}.\\ Here we will use the spectral dimension to probe the effective structure of space-time emerging in polymer field theory specializing to four space-time dimensions. Since in the IR, i.e. when all scales are much larger than the polymer scale, our theory reduces to ordinary field theory on Minkowski space we expect that that for large diffusion times ($s\rightarrow\infty$) the polymer spectral dimension will be just $d_s=4$. We will see that this is indeed the case but we will also study the running of $d_s$ at the polymer scale and below.\\ 
The Fourier transformed polymer Laplacian we will be considering will have a ``frequency" term and a non-trivial part determined by our ``one-particle" polymer dispersion relation (\ref{loc_en}) for $m=0$ \footnote{Here $\Delta_{\gamma}({\vec{p}})$ is the polymer correction to the energy of one-particle states introduced in Sec. IVB, not to be confused with the Laplace operator.}
\be
p^2= \omega^2 +(\Delta_{\gamma}({\vec{p}})|\vec{p}|)^2\,.
\ee
We start by considering the return probability
\be
P_{\gamma}(s) =\int\frac{d\omega}{2\pi}e^{-\omega^2 s}  \int\frac{d^3p}{(2\pi)^3}e^{- (\Delta_{\gamma}({\vec{p}})|\vec{p}|)^2 s} = P^{\omega}(s)P^{\vec{p}}_{\gamma}(s)\, .
\ee
The integral over the energy $P^{\omega}(s)$ is straightforward and returns just a constant contribution by 1 to the overall spectral dimension. We thus focus on the spatial integral 
\begin{equation}
P^{\vec{p}}_{\gamma}(s) = \int \frac{p^2 d p}{(2\pi)^3}e^{-(\Delta_{\gamma}({\vec{p}})|\vec{p}|)^2 s}.
\end{equation}
and evaluate its contribution to the spectral dimension using (\ref{spectr}). The integration can not be done analytically, therefore we performed a numerical evaluation and plotted the resulting polymer spectral dimension $d^{\gamma}_s=-2\frac{\partial \log P_{\gamma}(s) }{\partial\log s}$ in Figure \ref{pol_spectr} in units $\ell=1$ . 
As expected for large diffusion times the contribution to the spectral dimension of the spatial momentum integral $P^{\vec{p}}_{\gamma}(s)$ approaches the value of 3 i.e. we re-obtain the  flat space value of $d^{\gamma}_s = 4$ for the IR polymer spectral dimension. We see that as the diffusion time decreases the {\it spatial} spectral dimension increases up to a maximum value of $3.3$ around the scale $\ell=1$, where we thus have $d^{\gamma}_s\simeq 4.3$. At sub-polymer scales we see a fast decrease of the spectral dimension which in the limit $s\rightarrow 0$ again approaches the ordinary value $d^{\gamma}_s\simeq 4$. Thus the overall behaviour of the polymer spectral dimension is that of an overall {\it super-diffusion}, i.e. the spectral dimension is larger than the Hausdorff dimension at least in the interval $1\lesssim s\lesssim \mathcal{O}(100)$ moreover the value of $d^{\gamma}_s$ never decreases below the Hausdorff value of 4. This is a quite peculiar feature which distinguishes the polymer field theory from many other models beyond ordinary field theory motivated by Planck scale physics. In particular the characteristic {\it dimensional reduction} which appears to be a common denominator in all quantum gravity scenarios so far considered in the literature \cite{Ambjorn:2005db,Lauscher:2005qz,Horava:2009if,Benedetti:2008gu,Modesto:2008jz,Carlip:2009kf,Calcagni:2010pa,Sotiriou:2011mu,Alesci:2011cg,Amelino-Camelia:2013cfa,Arzano:2014jfa} is here absent. However our model shares some features like super-diffusion above the characteristic scale of the theory which have appeared in certain non-commutative models emerging in the semiclassical treatment of particles coupled to three-dimensional Einstein gravity \cite{Alesci:2011cg} and in certain approaches to causal dynamical triangulation \cite{Sotiriou:2011mu}.

\section{Discussion}
The main objective in this paper was to shed light on the effective space-time picture emerging in the polymer quantization of a scalar field. This unconventional quantization procedure introduces a fundamental polymer scale directly at the level of field space. Our work explored how this scale can affect field theoretic measurements which deal with scale-sensitive measurements or processes.\\
We first analyzed two alternative procedures for measuring energy in a localized region. In the first case we observed that the expectation value of a {\it local} polymer Hamiltonian on a polymer one-particle state is not sensitive to polymer corrections for what concerns the factor containing the information about the size of the local volume to which we are restricting the observable. Thus restricting the Hamiltonian to a local volume just produces an expectation value given by the polymer energy density times the local volume.  In the second case we considered a polymer localized one-particle state given by a superposition of global one-particle states modulated by a gaussian profile function whose width determined the extension of the region occupied by the quantum state. We found that also in this case polymer corrections affected only mildly the expectation value of the Hamiltonian i.e. the value of the energy associated to the state as a function of the width of the window function. In particular we found that it is still possible to ``squeeze" the localization region of the polymer state to zero while having a divergent expectation value of the energy in analogy with ordinary field theory. This indicates that, on one hand the uncertainty principle in this polymer field theoretic context is still affecting these types of localization measurements without significant changes below the polymer scale. On the other hand we see that, at least in this context, the polymer scale does not play the role of {\it minimal scale} nor it provides a cut-off scale capable to tame the short scale divergencies present in ordinary field theory.\\
This behaviour is somewhat reflected in the diffusion properties of the polymer corrected Laplacian which we considered in the last Section. Here, unlike many other examples of Planck-scale deformed or unconventional field theories, there is no running of the spectral dimension to values lower than the Haudorff dimension. Rather the spectral dimension remains larger than the Hausdorff dimension peaking around the polymer scale thus exhibiting {\it superdiffusion}. Below the polymer scale the spectral dimension quickly returns to the ordinary value of 4 indicating that the model does not exhibit exotic behaviours in the deep UV as confirmed by our previous analyses of localization.\\
Of course our results rely on the particular approach to polymer quantization based on a reformulation of Fock space which, while convenient to many extents, has some drawbacks and could be subject to improvement. A possibility would be to find an alternative candidate vacuum state which is annihilated by the associated destruction operator. As seen above such desirable property is absent in our approach. It would also very interesting to extend the localization analyses presented here in non-commutative theories in which a UV scale is introduced at the momentum space level through curvature of the geometry. Here it is known \cite{Alesci:2011cg,Amelino-Camelia:2013cfa,Arzano:2014jfa} that diffusion processes are very sensitive to the corrections introduced by such scale for small diffusion times. This will be the subject of upcoming work.

\section*{Acknowledgements}
We thank A. Kreienbuehl for useful correspondence. The work of MA is supported by the EU Marie Curie Actions through a Career Integration Grant and in part by a Grant from the John Templeton Foundation. ML thanks the ``Sapienza" University of Rome where the work leading to this paper was carried out.

\end{document}